\documentclass[twocolumn, twocolappendix]{aastex631}

\usepackage{amsmath}
\usepackage{graphicx}
\usepackage{orcidlink}

\usepackage{siunitx}
\DeclareSIUnit{\Lsun}{\text{\ensuremath{L_\odot}}}
\DeclareSIUnit{\Rsun}{\text{\ensuremath{R_\odot}}}
\DeclareSIUnit{\Msun}{\text{\ensuremath{M_\odot}}}

\newcommand{\sbs}[1]{\ensuremath{_\text{#1}}}

\begin{document}


\title{Collisional Evolution around A-type Stars with Strong Silicate Features}

\correspondingauthor{T. A. Stuber}
\email{tstuber@arizona.edu}

\author[0000-0003-2185-0525]{T. A. Stuber}
\affiliation{
    Department of Astronomy and Steward Observatory,
    The University of Arizona,
    933 North Cherry Ave,
    Tucson, AZ 85721, USA
}

\author[0000-0001-8808-2132]{H. Kobayashi}
\affiliation{
    Department of Physics,
    Nagoya University,
    Nagoya, Aichi 464-8602, Japan
}

\author[0000-0001-7841-3452]{S. Wolf}
\affiliation{
    Institute of Theoretical Physics and Astrophysics,
    Kiel University,
    Leibnizstr. 15,
    24118 Kiel, Germany
}

\author[0000-0001-6403-841X]{V. Faramaz-Gorka}
\affiliation{
    Department of Astronomy and Steward Observatory,
    The University of Arizona,
    933 North Cherry Ave,
    Tucson, AZ 85721, USA
}


\begin{abstract}
\noindent
Around luminous A-type stars, retaining micrometer-sized dust grains is
difficult owing to the high radiation pressure that repels particles with sizes
below the blowout limit. Yet such particles have been evidenced in the
terrestrial zone around those stars, via pronounced silicate spectral features.
An in situ planetesimal belt that collisionally replenishes the dust is
expected to deplete quickly, limiting the lifetime of observable features.
However, modeling of such systems often relies on a power-law description of
the grain size distribution that has a lower size limit owing to blowout. Using
more complex, collisional simulations instead, we investigated the temporal
evolution of hypothetical planetesimal belts in the terrestrial zone around
five A-type stars showing strong spectral silicate features. A collisional
cascade that is modified by radiation pressure results in dust distributions
that replicate observed spectral features without requiring other physical
processes. While features can be altered by particles below the blowout limit,
their magnitude is dominated by the emission of larger grains. Using the
systems' ages, we investigated the origin of the planetesimal belt. For the
three younger systems of our sample, a planetesimal belt that emerged from the
protoplanetary disk phase can explain the observations. This is even possible
for two older systems if the planetesimals are located at slightly larger
distances than the dust, e.g., if the dust drifted inward owing to the
Poynting–Robertson drag. Alternatively, a recent production of the
planetesimals, for instance, in a giant collision event, could have happened
only several million years ago.
\end{abstract}


\section{Introduction}

Debris disks are circumstellar distributions of dust, planetesimals, and trace
amounts of gas that were discovered by an infrared excess above stellar
photospheric emission, caused by the thermal emission of dust grains
\citep{aumann:1984}. While the dust grains get destroyed by multiple processes,
they are continuously replenished by collisions or evaporation of the
planetesimals that thus act as dust reservoirs \citep[for reviews, see,
e.g.,][]{krivov:2010, matthews:2014, hughes:2018}.

With the advent of the Spitzer Space Telescope \citep[hereafter
Spitzer;][]{werner:2004} and its Infrared Spectrograph
\citep[IRS;][]{houck:2004} and Multiband Imaging Photometer
\citep[MIPS;][]{rieke_GH:2004}, hundreds of main-sequence systems were found
to show a mid- to far-infrared excess over the stellar photosphere that was
best modeled jointly by a warm (with temperatures $\gtrsim \qty{100}{K}$) and
a cold ($\lesssim \qty{150}{K}$) dust distribution \citep{chen:2009, chen:2014,
morales:2011, ballering:2013}.  Most of these two-temperature debris disks
likely host two spatially distinct dust populations, especially in the case of
A-type stars \citep{kennedy:wyatt:2014}.

A subgroup of these disks shows spectral features in the mid-infrared that are
associated with silicates  \citep[e.g., ][and references in their
Sect. 1]{chen:2006, ballering:2014, mittal:2015}, most prominently around
wavelengths of $\lambda \sim \num{10}$ and \qty{18}{\um}. Such features
are interpreted as the emission of dust grains of at most several micrometers
in radius and at temperatures of several hundred kelvin, thus being located in
the terrestrial zone around the host stars ($\lesssim \qty{10}{au}$).

A possible origin for such warm dust distributions is a collisional cascade
in an in situ planetesimal belt \citep{kennedy:wyatt:2013}. However, as the
planetesimals are collisionally processed with time, the dust mass diminishes,
an effect that is mostly unaffected by the initial belt mass, as initially more
massive belts process faster \citep{dominik:decin:2003, wyatt:2007a,
kobayashi:tanaka:2010}. Thus, an in situ collisional cascade generally cannot
produce a given infrared excess at arbitrarily large times. The time when the
ability ceases increases along with the size of the largest planetesimals in
the belt, but there are physical limits to this, due to the planetesimal
formation processes \citep{kenyon:bromley:2002, kobayashi:2016}.
For silicate spectral features in observations of A-type star systems, another
challenge exists. The radiation pressure of these luminous host stars removes
dust grains below the blowout size out of the system \citep{burns:1979,
krivov:2006}. This process can remove grains comparable in size to those that
are needed to produce the silicate features. Frequently, spectral energy
distributions (SEDs) of warm dust distributions are modeled using a power-law
description of the grain size distribution that has a lower size limit
\citep[e.g.,][]{chen:2006, ballering:2014, mittal:2015}. However, this
simplified treatment of the size distribution has limitations. Generally, the
relative abundance of small grains modifies the size distribution for larger
grains \citep{campo_bagatin:1994}. In addition, assuming a lower cutoff that
is set by the blowout of small grains is only valid if the timescale on which
the blowout occurs is much smaller than the collisional timescale on which the
subblowout grains are replenished. In the limit of tenuous and optically thin
debris disks, the blowout timescale is independent of the dust number density,
while the collisional timescale decreases with increasing dust number density.
Thus, assuming a lower cutoff is disfavored for systems with a sufficiently
high dust number density where the impact of subblowout grains might no longer
be safely neglected. \citet{thebault:kral:2019} numerically showed the large
impact of subblowout grains on the size distribution and the resulting SEDs
for warm dust distributions around A-type stars. Observationally,
\citet{mittal:2015} found that, although the minimum grain size increases with
the stellar luminosity, this increase occurs at a lower rate than expected from
radiation pressure alone. \citet[][hereafter BA14]{ballering:2014} analyzed
SEDs of A-type systems that show spectral silicate features and found that the
required minimum grain sizes are on the same order, but often smaller than the
theoretically estimated blowout sizes.

An alternative explanation for the origin of the small and warm dust grains are
giant collisions of large objects, such as planetary embryos, that are
associated with the late stages of terrestrial planet formation
\citep[e.g.,][]{kenyon:bromley:2004, kenyon:bromley:2005, jackson:wyatt:2012,
genda:2015, wyatt:jackson:2016}. If an in situ planetesimal belt could be
excluded as the origin of spectral silicate features, the giant impact scenario
would be strengthened. Besides, the powerful space-based mid-infrared
observational window to study debris disks has just recently been opened again
by the Mid-Infrared Instrument \citep[MIRI;][]{rieke_GH:2015, wright:2023} with
its Medium-Resolution Spectrometer \citep[MRS;][]{wells:2015, argyriou:2023} on
board the James Webb Space Telescope \citep[JWST;][]{gardner:2006}, allowing
for new detailed studies of spectral silicate features
\citep[e.g.,][]{chen:2024, chai:2024, worthen:2024}.

In this study, we investigate the collisional evolution of warm debris disks in
the terrestrial zone around A-type stars that show strong spectral silicate
features. Using collisional simulations, we can avoid simplified power-law
treatments of the grain size distribution and test the hypothesis that the
collisional evolution of a planetesimal belt is able to explain the silicate
features as seen in observations. Based on those results, we use an analytical
description of quasi-steady-state collisional evolution to assess whether the
observed infrared excesses can be produced by a primordial planetesimal belt,
that is, a belt that has formed during the protoplanetary phase or shortly
after, or the planetesimal belt has to be of recent origin, for instance, was
produced in a recent major collision event.

Our collisional model will be described in Sect.~\ref{sect_col_model}, followed
by a presentation of our particular choice of parameters and simulation results
in Sect.~\ref{sect_col_evo_Atype}. The analysis of the planetesimal belt origin
is outlined in Sect.~\ref{sect_primordial_vs_recent}. After discussing our
assumptions in Sect.~\ref{sect_discussion}, we will conclude with a summary in
Sect.~\ref{sect_summary}.


\section{Collisional Model}
\label{sect_col_model}

We present our method to derive the collisional evolution of spherical bodies
in a narrow debris disk. The time evolution of the surface number density of
bodies with radii from $a$ to $a + da$, $n\sbs{s}(a) da$, is calculated via
\begin{eqnarray}
    \frac{\mathrm{d} n\sbs{s}(a)}{\mathrm{d} t}
    &=& \frac{1}{2} \int_{a\sbs{min}}^{a\sbs{max}}
        \int_{a\sbs{min}}^{a\sbs{max}} \mathrm{d} a_1 \mathrm{d} a_2
        n\sbs{s}(a_1) n\sbs{s}(a_2)
    \nonumber
    \\
    && \times K(a_1,a_2) F(a,a_1,a_2)
    \nonumber
    \\
    && - n\sbs{s}(a) \int_{a\sbs{min}}^{a\sbs{max}}
       \mathrm{d} a_2 n\sbs{s}(a_2) K(a,a_2)
    \nonumber
    \\
   && - \frac{n\sbs{s}(a)}{t\sbs{bo}(a)}
    \, ,
\end{eqnarray}
where $t$ is the time, $a_1$ and $a_2$ are the radii of the two colliding
bodies, $a\sbs{min}$ and $a\sbs{max}$ are the minimum and maximum radii of all
bodies, respectively, $K$ is the collisional kernel, $F$ is the collisional
outcome function that determines the size distribution of the collision
products, and $t\sbs{bo}$ is the blowout timescale of bodies due to radiation
pressure. The bin width is set to $\mathrm{d}a \approx 0.063 a$. 

The collisional kernel $K$ generally depends on the collisional velocities
\citep[e.g.,][]{kobayashi:2010b, kobayashi:hidekazu:21}. However, in debris
disks the collisional velocities are much higher than the mutual escape
velocities of the colliders. Thus, the collisional cross sections are given
by the geometrical cross sections, $\pi\cdot(a_1+a_2)^2$, and it follows that
\citep[see derivation in][]{kobayashi:tanaka:2010}
\begin{equation}
    K(a_1,a_2) = \xi \chi(a_1,a_2) \Omega \pi\cdot(a_1+a_2)^2 \, ,
    \label{eq_kernel}
\end{equation}
where $\xi = 0.88$ is the correction factor due to the three-body effect
caused by colliders revolving around the host star
\citep{greenzweig:lissauer:1992, kobayashi:tanaka:2010}, $\chi(a_1,a_2)$ is the
modification factor of the collision rate due to the increase in eccentricities
by radiation pressure (see below in this section), and
$\Omega = (G M_\star /R\sbs{disk}^3)^{1/2}$ is the Keplerian frequency, with
the gravitational constant $G$, the mass of the host star $M_\star$, and the
center radius of a disk $R\sbs{disk} =  \left(R\sbs{in}+R\sbs{out}\right)/2$
with inner/outer radius $R\sbs{in}/R\sbs{out}$. The collisional outcome is
controlled by the dimensionless impact energy
\begin{equation}
    \phi = \frac{E\sbs{imp}}{(m_1+m_2) Q\sbs{D}^\star} \, ,
\end{equation}
where $m_1$ and $m_2$ are the masses of the colliders, $E\sbs{imp}$ is the
impact energy, and $Q\sbs{D}^\star$ is the specific impact energy required for
the ejection of half the total mass of the colliders. According to
\citet{kobayashi:2010b}, we model the collisional outcome function as
\begin{equation}
    F(a,a_1,a_2) = \delta(a^3-a_1^3-a_2^3+a\sbs{e}^3) + F\sbs{eje} \, ,
    \label{eq_col_outcome}
\end{equation}
where $\delta(x)$ is the Kronecker-delta function and $a\sbs{e}$ is the radius
of a body with the mass of the total bodies ejected from a collision, given by
\begin{equation}
    a\sbs{e}^3 = \frac{\phi}{\phi+1} (a_1^3+a_2^3) \,.
\end{equation}
The first term on the right-hand side of eq.~\ref{eq_col_outcome} indicates the
largest remnant body, while the second term, $F\sbs{eje}$, represents the size
distribution of ejecta. The radius of the largest ejected particle $a\sbs{L}$
is defined as 
\begin{equation}
a\sbs{L}^3 = \frac{\epsilon \phi}{(1+\phi)^2} (a_1^3+a_2^3) 
\end{equation}
and the size distribution of ejecta is given by
\begin{equation}
    F\sbs{eje} = \frac{3(2-b)}{a} 
    \frac{1+\phi}{\epsilon} 
    \left( \frac{a}{a\sbs{L}} \right)^{3(1-b)}\textrm{, \,for } a \leq a\sbs{L}
\end{equation}
and $F\sbs{eje} = 0$ for $a > a\sbs{L}$. Conforming with
\citet{kobayashi:tanaka:2010}, we set $b=5/3$ and $\epsilon = 0.2$. 

Similar to previous studies \citep{benz:asphaug:1999, loehne:2008,
steward:leinhardt:2009, kobayashi:2010b}, we assume $Q\sbs{D}^\star$ to be
given by
\begin{eqnarray}
    \frac{Q\sbs{D}^\star}{\qty{1}{erg \,\g^{-1}}}
    &=& 2.0 \times 10^7 \left( \frac{a}{\qty{1}{\cm}}\right)^{-0.3}  
      + 1.1 \left( \frac{a}{\qty{1}{\cm}}\right)^{1.4}
      \nonumber
      \\
     && +\ 2.1 \times 10^{-5} \left( \frac{a}{\qty{1}{\cm}}\right)^{2},
    \label{eq_qds}
\end{eqnarray}
with $a$ referring to the radius of the larger collider. The factors of the
second and third terms are estimated assuming a bulk density of
$\rho = \qty{3.71}{\g\per\cubic\cm}$, suitable for amorphous olivine
\citep[MgFeSiO\sbs{4};][]{dorschner:1995}. For bodies with radii
$a \lesssim \qty{100}{m}$, the first term dominates and $Q\sbs{D}^\star$ is
determined by the material strength. The second term dominates for
$\qty{1}{km} \lesssim a \lesssim \qty{1000}{\km}$ and the third term dominates
for $a \gtrsim \qty{1000}{\km}$. While for both the second and third terms the
body's self-gravity dominates over the material strength, the described
regimes differ in the dominating mechanism to dissipate collisional energy
\citep{jutzi:2015, kobayashi:2026}. We assume the collisional velocity to be
\begin{equation}
    v = \sqrt{e^2+i^2} R\sbs{disk} \Omega
    \label{eq_col_velocity}
\end{equation}
where we set $e=2i=0.2$ comparable to the mean values of asteroid belt objects
in the Solar System \citep{ueda:2017}.

During the collisional evolution, bodies feel the specific radiation pressure
force $\beta GM_\star/R\sbs{disk}^2$ assuming geometrical optics, where $\beta$
is the ratio of the forces exerted by stellar radiation pressure and stellar
gravity \citep{zook:berg:1975}; particles with radii smaller than the blowout
radius $a\sbs{bo}$ are removed from the system (see
Appendix~\ref{appdx_sect_blowout_radius} for the derivation of $a\sbs{bo}$
using Mie theory to describe the radiation pressure force). In the collisional
simulations, we set $\beta \approx \beta^\prime = a \sbs{bo}/2 a$, which is a
good approximation for $a \gtrsim a\sbs{bo}$. We assume the average blowout
timescale to be the time required for dust to be blown across the disk
half-width $\Delta R\sbs{disk}/2 = (R\sbs{out}-R\sbs{in})/2$, given by
\begin{equation}
    t\sbs{bo} = \left(
    \frac{R\sbs{disk}^2 \Delta R\sbs{disk}}{\beta^\prime G M_*}
    \right)^{1/2}
\end{equation}
for $a \leq a\sbs{bo}$ and $t\sbs{bo} = \infty$ otherwise.

\begin{deluxetable*}{c c c c c c c c c c c c}
    \tabletypesize{\footnotesize}
    \setlength{\tabcolsep}{4pt}
    \tablecaption{
        Assumed Stellar Properties and the Resulting Blowout Grain
        Size
        \label{tab_belt_star_params}
    }
    \tablehead{
        \colhead{HIP Identifier} &
        \colhead{HD Identifier} & \colhead{Sp. Type} &
        \colhead{$t\sbs{age}$ /\unit{Myr}} &
        \colhead{$L_\star$ /\unit{\Lsun}} &
        \colhead{$T\sbs{eff}$ /\unit{K}} &
        \colhead{$R_\star$ /\unit{\Rsun}} &
        \colhead{$M_\star$ /\unit{\Msun}} &
        \colhead{$d_\star$ /\unit{pc}} &
        \colhead{$R\sbs{in}$ /\unit{au}} &
        \colhead{$R\sbs{out}$ /\unit{au}} &
        \colhead{$a\sbs{bo}$ /\unit{\um}} \\
    }
    \startdata
        HIP 41081 & HD \phn71043  & A0V    & \phn\num{30} & \num{28.86}    & \num{9800} & \num{1.66} & \num{2.90} & \phn\num{70}\phd\phn & \num{2.5} & \num{4.0} & \num{2.58} \\
        HIP 59394 & HD 105850     & A1V    & \num{369}    & \num{23.61}    & \num{9400} & \num{1.84} & \num{2.64} & \phn\num{58.5}       & \num{1.5} & \num{3.0} & \num{2.92} \\
        HIP 66068 & HD 117665     & A1/2V  & \phn\num{20} & \num{29.57}    & \num{9400} & \num{2.06} & \num{2.64} & \num{147.9}          & \num{4.5} & \num{5.0} & \num{3.62} \\
        HIP 78641 & HD 143675     & A5IV/V & \phn\num{20} & \phn\num{7.38} & \num{8190} & \num{1.35} & \num{2.00} & \num{113.4}          & \num{3.0} & \num{4.0} & \num{1.28} \\
        HIP 79797 & HD 145689     & A4V    & \num{203}    & \num{10.30}    & \num{8434} & \num{1.51} & \num{2.12} & \phn\num{52.2}       & \num{0.6} & \num{1.0} & \num{1.66} \\
    \enddata
\end{deluxetable*}

However, even if no blowout occurs for particles with $a > a\sbs{bo}$, the
radiation pressure still increases orbital eccentricities, which modifies the
collisional rate between particles \citep{thebault:wu:2008}. Considering
$a_1 \geq a_2$, we then incorporate this effect into eq.(\ref{eq_kernel}) using
$\chi(a_1,a_2)$, which is given by
\begin{eqnarray}
    \chi(a_1,a_2) = 
    \left\{ 
    \begin{array}{c l}
         \frac{\Delta R_{\rm disk} + \frac{\beta_1}{(1-\beta_1)}}{
               \Delta R_{\rm disk} + \frac{\beta_2}{(1-\beta_2)}}
         &\textrm{for} \; \beta_1 \leq \beta_2 < 0.5 \, , \\
         \\
         \frac{\Delta R_{\rm disk} + \frac{\beta_1}{(1-\beta_1)}}{
               \Delta R_{\rm disk} + 1}
         &\textrm{for} \; \beta_1 < 0.5 \leq \beta_2 \, ,
    \end{array}
    \right.
\end{eqnarray}
with the values $\beta_1$ and $\beta_2$ corresponding to the particle radii
$a_1$ and $a_2$. Here $\beta/(1 - \beta)$ is the orbital eccentricity of a
particle that is launched from a circular parent belt, while
$\Delta R_{\rm disk} + \beta/(1 - \beta)$ is the sum of the radial broadening
of a belt caused by the spread of semi-major axes and eccentricities.

Collisions of bodies with radii of \num{1}–\qty{100}{\km} surpass the critical
impact energy $Q\sbs{D}^\star$ once the collisional velocity is
$\sqrt{Q\sbs{D}^\star} \gtrsim$ \qty{30}{\m\per\s} to \qty{1}{\km\per\s}
(eq.~\ref{eq_qds}). Assuming that the considered systems are similar to the
solar system's asteroid belt with a typical collisional velocity of
$\sim \qty{5}{\km\per\s}$ \citep{bottke:1994}, the collisional velocity is much
larger than $\sqrt{Q\sbs{D}^\star}$. Thus, collisions induce fragmentation and
not growth, and the initially largest objects, the parent bodies with radius
$a\sbs{pb}$, remain the largest bodies throughout the simulation, and hence
$a\sbs{max} = a\sbs{pb}$. The collisional fragmentation, together with the
removal of small dust particles by radiation pressure, reduces the amount of
material in the disk over time.

The total amount of bodies is characterized by the total vertical geometrical
optical depth $\tau$ (hereafter total optical depth), given by 
\begin{eqnarray}
 \tau &=& \int_{a\sbs{min}}^{a\sbs{max}} \pi a^2 n\sbs{s}(a) \mathrm{d} a
 \nonumber
 \\  
 &=& \int_{a\sbs{min}}^{a\sbs{max}} \Delta \tau(a) \mathrm{d} \ln a \, ,
 \label{eq_tau}
\end{eqnarray}
where
\begin{equation}
    \Delta \tau(a) = \pi a^3 n\sbs{s}(a) \, .
    \label{eq_delta_tau}
\end{equation}

The total optical depth $\tau$ describes the total cross section of all
particles, while the size distribution of the optical depth $\Delta \tau(a)$
describes the cross section of particles with radii around $a$; both quantities
are given per unit area and are thus dimensionless. Simulations are initialized
with a total optical depth of $\tau = \tau_0$ and a size distribution of
$n\sbs{s}(a) \propto a^{-3.5}$ for bodies ranging from $a = a\sbs{bo}$ to
$a\sbs{pb}$ and $n\sbs{s}(a) = 0$ for $a < a\sbs{bo}$ and $a > a\sbs{pb}$. The
choice of \num{-3.5} for the power-law exponent is arbitrary, because this will
be erased during collisional evolution.

For small bodies with radii $a \sim a\sbs{bo}$, the collisional timescale is
comparable to $t\sbs{K}/\tau$, where $t\sbs{K}= 2\pi/\Omega$ is the orbital
timescale. Because this timescale is shorter than the simulated system age, the
size distribution for small bodies is always determined by that of the
collisional cascade in the quasi-steady-state. However, for larger bodies, the
collisional timescale is longer. Thus, the collisional cascade in the
quasi-steady-state controls the full size distribution only once the simulated
time exceeds the collisional timescale for bodies with $a = a\sbs{pb}$. Then,
the decrease of $\tau$ with time and hence the mass loss of the system is
described as \citep{kobayashi:tanaka:2010}
\begin{equation}
    \tau = \frac{\tau_0}{1+t/t\sbs{dep, 0}} \, ,
    \label{eq_mass_loss}
\end{equation}
with the initial depletion timescale $t\sbs{dep, 0}$ for $\tau \sim \tau_0$
(see derivation in Appendix~\ref{sect_depletion_timescale}) given by 
\begin{eqnarray}
    t\sbs{dep, 0} =
    1.5 \times 10^5 &&\left( \frac{a\sbs{pb}}{100\,{\rm km}}\right)^{0.97}
    \left( \frac{e\sbs{pb}}{0.2} \right)^{-0.68}
    \left( \frac{\tau_0}{10^{-2}} \right)^{-1}
    \nonumber
    \\ \times
    &&\left( \frac{a\sbs{bo}}{1\,\micron} \right)^{-0.63}
    \left( \frac{M_\star}{M_\sun} \right)^{-1.18}
    \nonumber
    \\  \times
    &&\left( \frac{R\sbs{disk}}{1\,{\rm au}}\right)^{2.18} \unit{yr} \, ,
\label{eq_dep_time}
\end{eqnarray}
where $e\sbs{pb}$ is the average orbital eccentricity of planetesimals
($e\sbs{pb} = 0.2$ in our simulations). For $t \gg t\sbs{dep, 0}$, the
depletion timescale of $\tau$ is given by
$t\sbs{dep} = \tau_0 t\sbs{dep, 0}/\tau$, which is independent of the initial
condition $\tau_0$. 


\section{Collisional evolution around A-type stars}
\label{sect_col_evo_Atype}


\subsection{System Parameters}
\label{subsect_parameters}

To model realistic systems, we use parameters as inferred from observations of
A-type stars. In particular, in Table~\ref{tab_belt_star_params} we collect
the parameters of five warm debris belts around stars with spectral types (Sp.
Type)\footnote{Obtained from the SIMBAD astronomical database
\citep{wenger:2000} based on \citet{houk:cowley:1975}, \citet{houk:1978,
houk:1982}, and \citet{houk:smith_moore:1988}.} between A0 and A5 as derived
by BA14. HIP 41081 (HD 71043), HIP 66068 (HD 117665), and HIP 78641 (HD 143675)
are young stars with estimated ages of $t\sbs{age} = \num{20}$–\qty{30}{Myr}
\citep{su:2006, rizzuto:2011}, while HIP 59394 (HD 105850) and HIP 79797 (HD
145689) are older, with ages of $t\sbs{age} \sim \qty{400}{Myr}$ and
${\sim}\qty{200}{Myr}$ \citep[see BA14, Sect. 2.1, using][]{schaller:1992,
zorec:royer:2012, nielsen:2013}.

Those systems show an infrared excess over the stellar photosphere as measured
by Spitzer/IRS. By modeling the excesses with two dust belts, BA14 revealed
pronounced spectral features produced by silicates. We use their spatial
constraints for the inner belt (see their Table~3) that is primarily
responsible for the silicate features. For the stellar properties we use the
values listed in their Table~1; those are based on the works of
\citet{cox:2000}, \citet{carroll:ostlie:2006}, and \citet{ballering:2013}.

To model the dust material, we follow BA14 and use the optical constants of
amorphous olivine \citep{dorschner:1995}\footnote{Obtained from
\url{https://www2.astro.uni-jena.de/Laboratory/OCDB/amsilicates.html}.} and the
same bulk density as for our collisional simulations (see
Sect.~\ref{sect_col_model}). We compute the blowout radii $a\sbs{bo}$ with the
procedure outlined in Appendix~\ref{appdx_sect_blowout_radius} and yield values
similar to those of BA14 (see Table~\ref{tab_belt_star_params}).

For the radius of the parent bodies we choose $a\sbs{pb} = \qty{100}{\km}$,
similar to the mass-weighted mean size of asteroid belt \citep{bottke:2005} and
Kuiper belt objects \citep{bernstein:2004}. As the ongoing collisional
evolution decreases $\tau$ and we compare our results with observations once the
collisional cascade is well established, the initial total optical depth
$\tau_0$ has to be sufficiently large. We find values of $\tau_0 = 10^{-2}$ for
HIP 66068 and HIP 78641 and $\tau_0 = 10^{-3}$ for the other systems to
suffice.


\subsection{Collisional Simulations}
\label{subsect_col_model}
 
\begin{figure}
    \centering
    \includegraphics[width=1\linewidth]{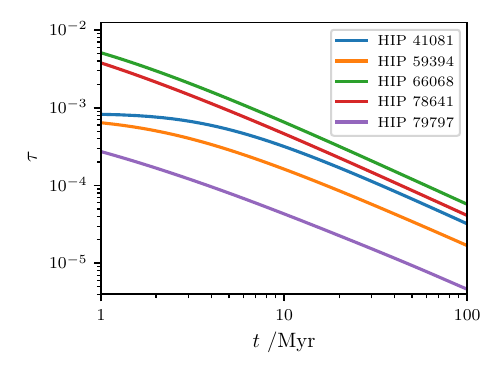}
    \caption{
        Time evolution of the total optical depth $\tau$ for the five simulated
        systems between \num{1} and \qty{100}{Myr}. For HIP 66068 and HIP
        78641, the simulations were started with an initial total optical depth
        of $\tau_0 = 10^{-2}$ while all other systems were initialized with
        $\tau_0 = 10^{-3}$ (see Sect.~\ref{sect_col_model}).
        \label{fig_total_opt_depth_over_time}
    }
\end{figure}
\begin{figure*}
    \centering
    \includegraphics[width=1\linewidth]{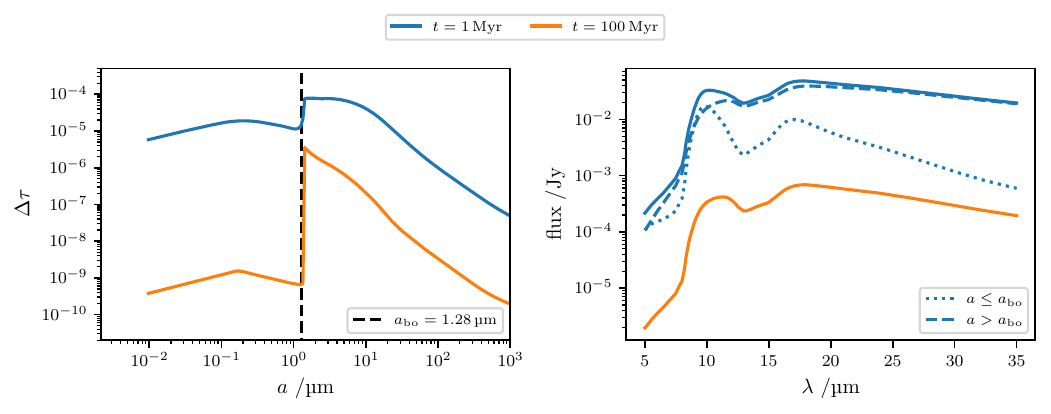}
    \caption{
        Simulation results for the system HIP 78641 for the times
        $t = \qty{1}{Myr}$ (blue) and $t = \qty{100}{Myr}$ (orange). Left: size
        distribution of optical depth $\Delta \tau$ for grains with radii $a$
        around the blowout grain radius $a\sbs{bo}$, which is indicated by the
        dashed black line. Right: corresponding SEDs in logarithmic scale. For
        $t = \qty{1}{Myr}$, the dotted line shows the contribution of grains
        with $a \leq a\sbs{bo}$ while the dashed line shows it for grains with
        $a > a\sbs{bo}$. For $t = \qty{100}{Myr}$, the contribution of
        particles below the blowout limit is negligible.
        \label{fig_opt_depth_sed}
    }
\end{figure*}

Figure~\ref{fig_total_opt_depth_over_time} shows the time evolution of the
total optical depth (eq.~\ref{eq_tau}). Once the collisional cascade is well
established across all size bins, it decreases the number of bodies without
modulating the shape of the size distribution for $a \gg a\sbs{bo}$. We confirm
that such size distributions are well achieved after ${\sim} \qty{1}{Myr}$. For
increasing time, the collisional cascade transfers mass from the largest bodies
down to smaller particles until the smallest particles are blown out of the
system by radiation pressure. Thereby, the systems lose mass and the total
optical depth decreases over time following eq.~\ref{eq_mass_loss}. In the
limit of $t \gg t\sbs{dep, 0}$, it holds that $\tau \propto 1/t$, which can be
seen in Fig.~\ref{fig_total_opt_depth_over_time} for $t \gtrsim \qty{10}{Myr}$.

The left panel of Figure~\ref{fig_opt_depth_sed} shows for the exemplary system
HIP 78641 the size distribution of the optical depth (see
eq.~\ref{eq_delta_tau}) for grain radii between $a = \qty{0.1}{\um}$ and
\qty{1}{mm} at times of $t = \num{1}$ and \qty{100}{Myr} (see
Fig.~\ref{fig_multi_opt_depth_sed} for all other systems). For an ideal
steady-state collisional cascade, this distribution is scale independent and
given by a power law \citep{dohnanyi:1969, tanaka:96, kobayashi:tanaka:2010},
and thus the total mass and thereby total optical depth only act as a scaling
factor. Particles with radii $\gtrsim \qty{10}{\um}$ that are sufficiently
larger than the blowout radius follow this behavior. Contrarily, for grain
radii close to and below the blowout radius, radiation pressure modifies the
distribution. Prominently, the optical depth for particles with radii below the
blowout radius is reduced, leading to the cutoff at $a = a\sbs{bo}$. The
magnitude of the decrease depends on the blowout time scale on which particles
are removed compared to the collisional time scale on which particles are
replenished. While the blowout time scale is independent of the total optical
depth $\tau$, the collisional time scale is shorter for higher values of
$\tau$. Therefore, the deficit of subblowout grains compared to larger grains
is smaller for earlier times and increases as the mass and total optical depth
decrease.

The wavy pattern for radii slightly larger than the blowout radius is a
consequence of the relative underabundance of subblowout particles
\citep{campo_bagatin:1994} and has been seen in previous numerical studies
\citep{thebault:2003, krivov:2006, loehne:2008, thebault:wu:2008,
thebault:kral:2019}. The shape of the size distribution of optical depth for
such particles directly above the blowout radius is generally dependent on
their abundance relative to subblowout particles and thus evolves with time.


\subsection{Observational Simulations}

To compute SEDs, we use the numerical tool \texttt{DMS} \citep{kim:2018} and
adjust it to use the size distributions from our collisional simulations.
\texttt{DMS} utilizes a wedge-like structure of which we choose the
half-opening angle to be \qty{5}{\degree}, resembling an almost flat ring. The
radial restrictions of the ring are identical to the collisional simulations
(Sect.~\ref{sect_col_model}). The inclination of the ring with respect to the
line of sight is set to be \qty{0}{\degree} (face-on); however, this has a
negligible impact on the simulation results, as the radiation emanating from
the dust grains is dominated by isotropic thermal emission at the investigated
wavelengths.

Based on the time evolution of the total optical depth $\tau$, we compute the
time evolution of the corresponding SEDs for each system. The right panel of
Figure~\ref{fig_opt_depth_sed} shows the SED corresponding to the size
distribution of optical depth $\Delta\tau$ on the left for the system HIP 78641
(see Fig.~\ref{fig_multi_opt_depth_sed} for all other systems). The later SED
is shifted toward lower fluxes while mostly resembling the shape of the earlier
SED, caused by the mostly shape-retaining decrease in optical depth for
particle radii $>a\sbs{bo}$. However, the varying relative abundance of
subblowout particles compared to larger ones slightly changes the shape of the
SED, most prominently seen around the silicate features at wavelengths of
$\lambda \sim \num{10}$ and \qty{18}{\um}. The flux contribution of particles
below $(a \leq a\sbs{bo})$ and above $(a > a\sbs{bo})$ the blowout limit are
shown for $t = \qty{1}{Myr}$; for $t = \qty{100}{Myr}$, the contribution of
subblowout particles is negligible.


\subsection{Relative excess of silicate spectral features}
\label{subsect_rel_excess}

\begin{figure*}
    \centering
    \includegraphics[width=1\linewidth]{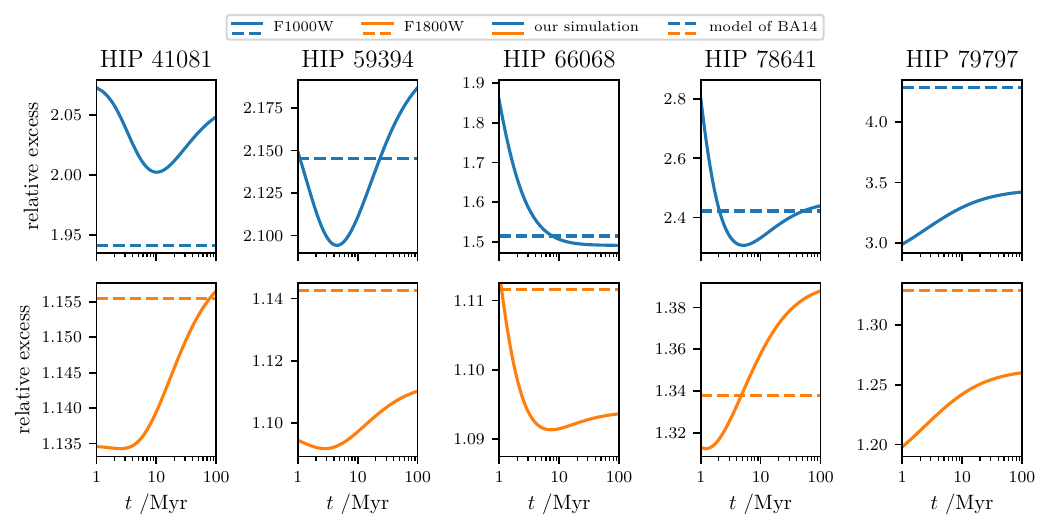}
    \caption{
        Time evolution of the relative excess caused by silicate spectral
        features around wavelengths of \qty{10}{\um} in the JWST/MIRI filter
        F1000W (blue) and around \qty{18}{\um} in the filter F1800W (orange).
        Dashed lines indicate the relative excess from the BA14 models.
    \label{fig_rel_excess_sil_feature}
    }
\end{figure*}

Motivated by the observational capabilities of JWST/MIRI, we define a measure
of the relative excess caused by silicates that can be used to interpret both
spectroscopic and broadband information of warm debris disks. We compute the
relative excess observed through an MIRI broadband filter covering a given
feature by the quotient of the flux through that filter and a reference flux.
The reference flux is defined as the flux at the center wavelength of that
filter, computed by linearly interpolating between the fluxes through filters
that bracket the silicate feature at shorter and longer wavelengths. The
wavelengths used for interpolation are the center wavelengths of the respective
filters. Effectively, the reference flux is an estimate of the flux that would
be measured if no silicate feature were present.

Using information about the filters from \citet{miri_user_documentation:2024},
we compute the relative excess of the silicate feature at
${\sim} \qty{10}{\um}$ using the filter F1000W (center wavelength
$\lambda\sbs{c} = \qty{10.0}{\um}$, filter width
$\Delta\lambda = \qty{1.80}{\um}$) and bracket it by the filters F770W
($\lambda\sbs{c} = \qty{7.7}{\um}$, $\Delta\lambda = \qty{1.95}{\um}$) and
F1500W ($\lambda\sbs{c} = \qty{15.0}{\um}$,
$\Delta\lambda = \qty{2.92}{\um}$). For the feature at ${\sim} \qty{18}{\um}$,
we use the filter F1800W ($\lambda\sbs{c} = \qty{18.0}{\um}$,
$\Delta\lambda = \qty{2.95}{\um}$) and bracket it by the filters F1500W and
F2100W ($\lambda\sbs{c} = \qty{21.0}{\um}$,
$\Delta\lambda = \qty{4.58}{\um}$). We assume all filters to have
\qty{100}{\percent} transmission over the whole bandpass; if this method is
used to interpret observations with JWST/MIRI, the individual transmission
curves of the filter might be used.

The temporal evolution of the relative excess
(Fig.~\ref{fig_rel_excess_sil_feature}) follows a general trend across systems
and filters that is best visible for HIP 78641 and F1000W (top row, fourth
panel). For early times, the relative excess decreases owing to the decreasing
relative abundance of subblowout grains compared to larger grains that directly
reduces the peak magnitudes (see Fig.~\ref{fig_opt_depth_sed}). After this
effect has ceased mostly, the grain size distribution still undergoes changes
for sizes slightly above the blowout size. This reduces flux in the areas
surrounding the peaks, that is, through the bracketing filters, which increases
the relative excess again. Lastly, this process also ceases, and the relative
excess approaches a constant value. Due to different simulation parameters,
each system experiences this evolution at different times and with a different
magnitude. Thus, for the shown fixed time interval, different sections and
scales of the effect are visible. For all systems, the relative excess of the
${\sim} \qty{10}{\um}$ feature is stronger than that of the
${\sim} \qty{18}{\um}$ feature. In addition, it shows larger variations in
time, the reason being that the relative shape of the grain size distribution
undergoes the most drastic changes for small grains, which impacts the SED
stronger at shorter wavelengths.

The relative excess for one or more features at a given time can be compared to
observational measurements to constrain both the dust material and the
collisional evolution of the system. For systems HIP 41081, HIP 59394, and HIP
78641, using only the \qty{10}{\um} feature gives degenerate results as the
same relative excess can be found at two different times within the considered
time interval. However, by additionally using the \qty{18}{\um} feature, or if
the age of the system is known sufficiently precisely, the degeneracy can be
broken.

From our simulations, we obtain excesses due to silicate features on a similar
magnitude compared to the models of BA14 (shown with dashed lines in
Fig.~\ref{fig_rel_excess_sil_feature}); HIP 79797 is the exception, with by far
the largest relative and absolute differences. These similarities, or the lack
thereof, will be further investigated in the next section.


\subsection{Reproducing Spitzer Observations}
\label{subsect_reproduce}

\begin{figure*}
    \centering
    \includegraphics[width=1\linewidth]{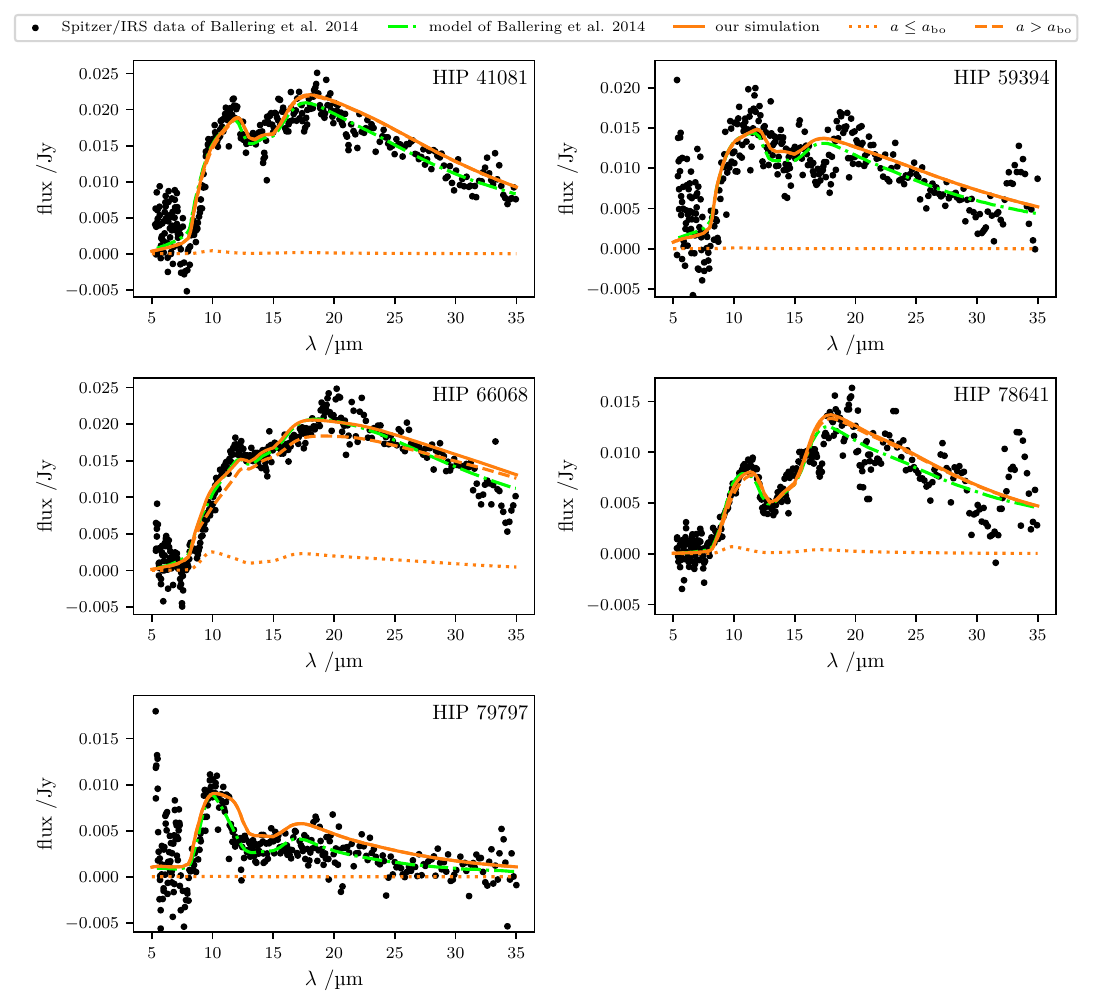}
    \caption{
        Comparison of our simulated SEDs (solid orange line) with the model
        results of BA14 (dash–dotted green line) for selected times with the
        magnitude of the flux peak around ${\sim} \qty{10}{\um}$ as the
        matching criterion. Orange dotted lines show the relative contribution
        of grains with $a \leq a\sbs{bo}$ in our simulations, while orange
        dashed lines show it for grains with $a > a\sbs{bo}$. Black circles
        denote Spitzer/IRS data as derived by BA14 by subtracting the flux
        contribution of the central star and a second, cold dust distribution.
        Some negative values resulting from the subtraction are omitted for
        better visual clarity.
        \label{fig_SED_comparison}
    }
\end{figure*}

We qualitatively select for each system the times for which the system's SED
resembles closely the modeling results from BA14 (see their Fig. 2, right
panel)\footnote{Models and data points from observations retrieved from the
article of BA14 with the online version
(\href{https://apps.automeris.io/wpd4/}{https://apps.automeris.io/wpd4/}) of
the WebPlotDigitizer, release v4.7, developed on GitHub
(\href{https://github.com/automeris-io/WebPlotDigitizer}{
https://github.com/automeris-io/WebPlotDigitizer}).} and show the matches in
Fig.~\ref{fig_SED_comparison}. As the matching criterion we choose the
magnitude of the flux peak around ${\sim} \qty{10}{\um}$. Our selected SEDs
are not a fit to the observations but result from collisional simulations using
the modeling results of BA14 (see Sect.~\ref{subsect_parameters}). For the
selected SEDs, the emission of subblowout grains is only relevant for HIP
66068 and HIP 78641 and negligible for other systems.

Across the systems, our simulations show SEDs with silicate features around
\qty{10}{\um} and \qty{18}{\um} similar in shape and magnitude to those in both
the Spitzer data and modeling results of BA14. Differences between our results
and the modeling results of BA14 arise in the exact shape of the silicate
features and in the slope and absolute values of the flux at longer wavelengths
${\gtrsim} \qty{17}{\um}$. This is a result from the two different approaches
(see Sect.~\ref{subsect_disc_matching_criterion}) and does not affect the
following conclusion: An in situ planetesimal belt undergoing collisional
evolution is able to explain pronounced silicate spectral features in
observations of the tested A-type star systems despite the strong radiation
pressure that removes small particles. Dependent on the system parameters, dust
grains below the blowout size can modify the shape and prominence of the
silicate spectral features, while grains above that limit have the dominating
effect on the SED.


\section{Primordial Planetesimal Belts versus Recent Formation}
\label{sect_primordial_vs_recent}

The question remains whether a primordial planetesimal belt formed during the
protoplanetary disk phase or shortly after can explain the observed infrared
excesses at the ages of the systems or would be processed too quickly, and
hence the planetesimals would have to be created or replenished at later times.

The times $t\sbs{match}$ for which our simulations match the modeling results
of BA14 best (Sect.~\ref{subsect_reproduce}) are listed in
Table~\ref{tab_estimate_params}. These times are earlier than the system ages
(see Table~\ref{tab_belt_star_params}; values are discussed later in this
section and Sect.~\ref{subsect_disc_sim_parameters}) but depend on our choice
of $a\sbs{pb} = \qty{100}{km}$ and $e\sbs{pb} = 0.2$ provided that
$R\sbs{disk}$ and $M_\star$ are good estimates. Assuming that a
quasi-steady-state collisional evolution began with the formation of the
systems at $t = \qty{0}{yr}$, the values of $a\sbs{pb}$ and $e\sbs{pb}$ that
are required to reproduce the SEDs at the system ages can be estimated using
the analytic description of quasi-steady-state evolution.

First, using eq.~\ref{eq_mass_loss} with $t=t\sbs{match}$, we estimate the
total optical depth $\tau\sbs{match}$ for which the SEDs are best reproduced
(see Table~\ref{tab_estimate_params}). These values differ slightly from the
values shown in Fig.~\ref{fig_total_opt_depth_over_time} for $t=t\sbs{match}$
as eq.~\ref{eq_mass_loss} assumes a quasi-steady-state throughout, while the
simulations have to settle into this first. Then, we compute, again with
eq.~\ref{eq_mass_loss}, the depletion time $t\sbs{dep, 0, match}$ necessary to
achieve $\tau\sbs{match}$ at the age of the systems, that is,
$t\sbs{match} = t\sbs{age}$.

In eq.~\ref{eq_dep_time}, $a\sbs{pb}$ and $e\sbs{pb}$ are degenerate in their
effect on $t\sbs{dep, 0}$. However, for collisions to be effective, the
collisional velocity (eq.~\ref{eq_col_velocity}) has to be sufficiently large
to disrupt the bodies, and thus $v \ga \sqrt{Q\sbs{D}^\star}$. We estimate the
minimum eccentricity necessary for collisions to disrupt parent bodies by
adopting $v = \sqrt{Q\sbs{D}^\star}$, and thus yield
\begin{equation}
    e\sbs{pb, min} = \sqrt{
        \frac{4}{5}\frac{Q\sbs{D}^\star R\sbs{disk}}{GM_\star}
    } \, .
    \label{eq_e_pb_min}
\end{equation}
We assume $Q\sbs{D}^\star$ to be given by the second term
of eq.~\ref{eq_qds}, suitable for planetesimals, which depends on $a\sbs{pb}$.
Then, eq.~\ref{eq_dep_time} with $t\sbs{dep, 0} = t\sbs{dep, 0, match}$ and
$e\sbs{pb} = e\sbs{pb, min}$ yields the minimal parent body radius $a\sbs{pb, min}$
that is required to reproduce the SEDs at the system ages while allowing for
collisional disruption of the parent bodies; the corresponding value
of $e\sbs{pb, min}$ is given by eq.~\ref{eq_e_pb_min} (see
Table~\ref{tab_estimate_params}).

\begin{deluxetable}{c c c c c}
    \tabletypesize{\small}
    \setlength{\tabcolsep}{4pt}
    \tablecaption{
        Match Times and Corresponding Total Optical Depth Based on Which the
        Minimum Parent Body Size and Average Eccentricity Required to Reproduce
        the SEDs at System Ages Are estimated
        \label{tab_estimate_params}
    }
    \tablehead{
        \colhead{HIP Identifier} &
        \colhead{$t\sbs{match}$ /\unit{Myr}} &
        \colhead{$\tau\sbs{match}$} &
        \colhead{$a\sbs{pb, min}$ /\unit{\km}} &
        \colhead{$e\sbs{pb, min}$}
    }
    \startdata
        HIP 41081 & \num{22}\phantom{.2} & \num{1.2e-4} & \phn\phn\num{10} & \num{0.006}       \\
        HIP 59394 & \num{31}\phantom{.2} & \num{4.4e-5} & \phn\num{790}    & \num{0.1}\phn\phn \\
        HIP 66068 & \phn\num{2.7}        & \num{1.9e-3} & \phn\num{480}    & \num{0.1}\phn\phn \\
        HIP 78641 & \phn\num{5.2}        & \num{2.1e-3} & \phn\num{170}    & \num{0.06}\phn    \\
        HIP 79797 & \phn\num{5.2}        & \num{7.4e-5} & \num{6200}       & \num{0.3}\phn\phn \\
    \enddata
    \tablecomments{
        Values of $a\sbs{pb, min}$ are rounded to tens of kilometers.
    }
\end{deluxetable}

For HIP 41081 and HIP 78641, the values of of $a\sbs{pb, min}$ and
$e\sbs{pb, min}$ are well possible, and thus a primordial planetesimal belt
can explain the observed SEDs. For HIP 66068, rather large parent bodies are
required. Its age of ${\sim} \qty{20}{Myr}$, however, is not precisely known.
In addition, there are uncertainties on when the collisional cascade is
established, while we assumed this to be the case ab initio, an issue that is
related to the planet formation process and dynamical stirring of the
planetesimal population \citep[see, e.g.,][]{kenyon:bromley:2001, wyatt:2008,
kennedy:wyatt:2010, kobayashi:2014,krivov:booth:2018}. If collisions are
delayed by \qty{10}{Myr} for HIP 66068, the values from
Table~\ref{tab_estimate_params} reduce to
$a\sbs{pb, min} \approx \qty{120}{\km}$, $e\sbs{pb, min} = \num{0.04}$. Thus,
we conclude that a primordial planetesimal belt can explain the observed SEDs
for HIP 66068 as well.

In the case of the ${\sim} \qty{200}{Myr}$ old system HIP 79797, small
uncertainties in the age estimation are of minor effect. Furthermore, the
estimated value of $a\sbs{pb, min} \approx \qty{6200}{\km}$ is underestimated
owing to the strong self-gravity of such objects \citep{steward:leinhardt:2009,
kobayashi:2026} as described by the neglected third term of eq.~\ref{eq_qds}.
Thus, at least Earth-sized bodies would have to undergo continuous collisions
at an eccentricity of ${\approx} \num{0.3}$ at a stellocentric distance of
$\sim \qty{1}{au}$, which can be excluded based on the grounds of structure
formation as follows. During the buildup phase in protoplanetary disks,
planetesimals can grow via collisions until runaway growth begins. Runaway
growth produces a small number of both protoplanets with wide orbital
separations and larger planetesimals. However, the majority of planetesimals no
longer grow beyond a size determined by the balance between turbulent stirring
and collisional damping \citep{kobayashi:2016}, and thus the characteristic
size of planetesimals is mainly determined by such smaller, abundant
planetesimals. Then, the number of smaller planetesimals is reduced by
collisional fragmentation due to stirring by protoplanets, which is less
efficient for larger planetesimals. This effect increases the characteristic
size of planetesimals by a factor of ${\sim} \num{5}$ \citep{kobayashi:2010b}.
Nonetheless, for a realistic turbulence strength, the characteristic
planetesimal radius is much smaller than \qty{3000}{km}. Thus, a primordial
planetesimal belt in steady state at the assumed location of the dust ring can
be excluded as the origin of the warm dust distribution for HIP 79797.

The oldest system in our sample, HIP 59394, is a mixed case. A sufficient
number of planetesimals with radii of at least ${\gtrsim} \qty{800}{\km}$ (more
than three times the size of Ceres) would have to participate in ongoing
collisions. In addition, the average parent body eccentricity $e\sbs{pb}$ might
be larger than our assumed minimal values, which would require even larger
planetesimals. However, the timescale on which dust particles slightly above
the blowout limit would fall onto the star under the Poynting–Robertson effect
\citep[e.g.,][]{burns:1979} is comparable to the collisional timescale on which
they would be destroyed \citep[][see also
Sect.~\ref{subsect_disc_sim_parameters}]{wyatt:2005}. Thus, the center radius
of the dust distribution might be located at smaller distances from the star
than the planetesimal belt. If we scale the center radius of the planetesimal
belt $R\sbs{disk}$ by a factor of \num{1.3}, for instance, if $R\sbs{disk}$ is
close to the outer radius of the dust distribution $R\sbs{out}$, the minimum
planetesimal radius shrinks to $a\sbs{pb, min} \approx \qty{300}{km}$ (slightly
larger than the asteroids 2 Pallas and 4 Vesta).

For HIP 79797, the belt radius would have to be underestimated by a factor of
${\approx} \num{2.1}$ to shrink $a\sbs{pb, min}$ to ${\approx} \qty{400}{km}$.
This would cause the planetesimal belt and the inferred location of the dust
distribution to be separated. While the dust distribution is not expected to be
severed from the belt, as the dust would be replenished in the planetesimal
belt and dragged from there, observational modeling might still infer the dust
at a distinctively different location. The emission power of the dust rises
with temperature and thus shrinking radius. Therefore, the mid-infrared flux
per grain peaks close to the inner edge of the dust distribution, where the
temperatures are highest. This can produce two rings in observations, one where
the dust number density is high and one where the emission power is high. See,
for instance, the bottom row of Figures 13 and 14 of \citet{loehne:2017}, where
two distinct rings emerge in maps of surface brightness that originate from a
single planetesimal belt.

Thus, we conclude that a primordial belt could be at the origin of the warm
dust around the ${\sim} \qty{400}{Myr}$ old system of HIP 59394 and the
${\sim} \qty{200}{Myr}$ old system of HIP 79797 if the dust gets dragged in
sufficient numbers by the PR effect. In the case of HIP 66068, an
underestimated radius of the planetesimal belt provides an alternative means of
shrinking $a\sbs{pb, min}$ to reasonable values besides the discussed delay of
collisions.

If, in contrast, a recent major collisional event has produced the planetesimal
belts in the older systems, our selected simulation parameters provide an
estimate of when this collision happened. Assuming that a population of
planetesimals with $a\sbs{pb} = \qty{100}{\km}$ and $e\sbs{pb} = \num{0.2}$ was
produced, this event happened ${\approx} \qty{30}{Myr}$ ago in the case of HIP
59394 and ${\approx} \qty{5}{Myr}$ ago in the case of HIP 79797 (see
Table~\ref{tab_estimate_params}).

Similar results were yielded by \citet{wyatt:2007a} for solar-like stars, who,
contrary to our approach, described the grain size distribution by a power law.
However, for their less luminous stellar sample, the blowout radius is smaller
if even present, and thus the effect of this assumption on the SED is small.


\section{Discussion}
\label{sect_discussion}


\subsection{Simulation Parameters}
\label{subsect_disc_sim_parameters}

We assumed a priori that the considered warm dust distributions are produced
by an in situ planetesimal belt and are not fed by an outer planetesimal belt
(a Kuiper belt analog) at tens of astronomical units with the dust migrating
inward under the PR effect. This is directly based on the modeling results of
BA14 that favor two distinct dust distributions, although we note that this
might be explained by a single parent belt as discussed in
Sect.~\ref{sect_primordial_vs_recent}. However, our assumption is supported by
\citet{kennedy:wyatt:2014}, who showed that SEDs showing two emitters at
different temperatures are likely caused by two spatially distinct dust
distributions. However, the warm dust distribution and its terrestrial parent
planetesimal belt might not be (completely) congruent, if the PR effect drags
dust grains inward and their survival against collisions is long enough to
shift the dust's radial center away from the parent planetesimal belt. In this
case, the distance of the simulated planetesimal belts from the star would be
larger than the assumed values. Whether this occurs can be estimated by
comparing the time needed for a dust grain at the blowout limit to fall into
the star against the collisional timescale \citep{wyatt:2005}. These timescales
are comparable for HIP 41081, HIP 59364, and HIP 79797, and thus we might have
underestimated the radius of their planetesimal belts. Future studies aiming
at characterizing particular systems should include PR drag in their
simulations \citep[e.g.,][]{loehne:2017, rigley:wyatt:2020, sommer:2025}.

Regarding the dust material, we chose amorphous olivine with a fixed bulk
density following BA14. The true dust composition is likely much more complex,
possibly blending different amorphous and crystalline silicates, carbonaceous
materials, and more \citep[e.g.,][]{lisse:2008, lisse:2012}. Furthermore, if
the dust grains are porous, this changes both the average bulk density of
grains and their optical properties, changing both the blowout radius and the
absorption and emission characteristics \citep{kirchschlager:wolf:2013,
brunngraeber:2017, stuber:wolf:2022}. Specifically, different levels of
porosity modify the contrast of the spectral feature in the absorption
efficiency of silicates: while the \qty{10}{\um} feature increases in contrast
for increasing porosity the \qty{18}{\um} feature increases for low porosities,
but then decreases for higher ones \citep[see Fig. 3
of][]{kirchschlager:wolf:2013}. This would lead to changes in the relative
excess due to the spectral silicate features (Sect.~\ref{subsect_rel_excess}).
Generally, a different dust composition means a different average bulk density
of the grains, which directly translates into shifts of the $\beta$-ratio and
thus blowout radius (see Appendix~\ref{appdx_sect_blowout_radius}). For our
comparison with Spitzer observations, the influence of subblowout grains is
negligible except for HIP 66068 and HIP 78641, while still being small, and
thus the effect of such bias in the blowout radius is limited. However, its
effect would increase for earlier times, when the relative abundance of such
grains is higher (see Fig.~\ref{fig_opt_depth_sed}).

Our estimates of $t\sbs{dep, 0}$ and $a\sbs{pb, min}$ via eq.~\ref{eq_dep_time}
depend primarily on the second term of $Q\sbs{D}^\star$ (see
eq.~\ref{eq_tdep_app}), which depends on the material properties of
planetesimals that are poorly known. In our assumption for this term, we used
values similar to the simulation results of \citet{benz:asphaug:1999} for
basalt and a collisional velocity of \qty{3}{\km\per\s} ($B$ in their Table~3;
see also their eq.~6). If we were to use their results for a collisional
velocity of \qty{5}{\km\per\s}, this would increase QD* by
${\approx} \qty{70}{\percent}$, which in turn would increase $t\sbs{dep, 0}$ by
a factor of ${\approx} \num{1.4}$ and $a\sbs{pb, min}$ by a factor of
${\approx} \num{2.0}$. Combined with the uncertainty on the planetesimals' bulk
density, we estimate that our results on $t\sbs{dep, 0}$ and $a\sbs{pb, min}$
are accurate to within a factor of a few.

In our assessment of the feasibility that a primordial planetesimal belt can
be the origin of the inferred warm dust distributions, we relied on the
stellar ages as compiled by BA14. For the young systems, the age when the
collisional cascade is established modifies this age estimate (see
Sect.~\ref{sect_primordial_vs_recent}). For the two older systems, BA14 note
that their used age estimates might have large uncertainties of
${\gtrsim} \qty{50}{\%}$. While we assumed the age estimates to be precise for
our assessment, we note that our conclusions might change if the systems are of
substantially different ages. The younger the system, the less strict are the
requirements on the system parameters to explain the warm dust being fed by a
primordial planetesimal belt.


\subsection{SED Matching Criterion}
\label{subsect_disc_matching_criterion}

Using the flux peak around \qty{10}{\um} to match the SEDs of our simulations
to the models of BA14 can lead to differing fluxes at longer wavelengths. These
deviations can be broadly split into two kinds. First, there can be an
approximately constant offset between fluxes in the two SEDs (e.g., HIP 79797)
caused by a different magnitude of the peak around \qty{10}{\um}. The model
approach of BA14 allowed for minimum grain radii $a\sbs{min}$ smaller than
$a\sbs{bo}$ and all of their models for the selected systems show
$a\sbs{min} < a\sbs{bo}$ with the largest relative difference arising for HIP
79797.

Second, there can be a differing slope of the SED at wavelengths
${\gtrsim} \qty{10}{\um}$ (e.g., HIP 66068). This can be explained by different
relative abundances of grains with different radii; smaller grains become less
efficient in emitting radiation at longer wavelengths (see, e.g., the different
slopes of the emission of subblowout and larger grains for
$\lambda \gtrsim \qty{17}{\um}$ in Fig.~\ref{fig_opt_depth_sed} and
\ref{fig_multi_opt_depth_sed}, right panel). While our size distributions of
optical depth (termed grain size distribution in BA14) had a fixed edge at the
blowout radius and an individual shape for smaller and larger dust grains as
determined by the collisional simulations (see Fig.~\ref{fig_opt_depth_sed} and
\ref{fig_multi_opt_depth_sed}, left panel), BA14 used a power law for the size
distribution with a fixed slope that stopped at a minimum grain radius
$a\sbs{min}$ determined by their fit.

Notwithstanding, the SED slope at these longer wavelengths deduced from
observations is usually impacted by the precision of the prior subtraction of
both the stellar photospheric emission and, more importantly, the emission of a
cold debris disk located at tens of astronomical units. In contrast, the shape
of the emission peaks is mostly unaffected by that.


\section{Summary}
\label{sect_summary}

We selected five A-type stars that show signs of a warm dust population and
strong ${\sim} \qty{10}{\um}$ and ${\sim} \qty{18}{\um}$ spectral silicate
features. Based on literature results from SED modeling, we performed
collisional and observational simulations of planetesimal belts in the
terrestrial zone around the stars as the assumed origin of the dust.

We showed that, despite the high radiation pressure environment that removes
dust grains of several micrometers in size, a collisional cascade started by
sufficiently dynamically excited planetesimals alone is able to produce the
spectral features and no further mechanisms are required. While the shape of
the spectral features can be modified by dust grains below the blowout limit,
they are dominantly produced by grains above this limit.

In addition, we tested whether a primordial planetesimal belt that emerged from
the protoplanetary disk phase and is undergoing collisional evolution since
then is able to explain the observed flux levels of the warm dust distributions
at the ages of the systems, assuming that planetesimals and dust are colocated.
For the three young systems with estimated ages of ${\sim} \num{20}$ and
\qty{30}{Myr}, this is a viable explanation.

For one older system with an estimated age of ${\sim} \qty{400}{Myr}$, this is
an unlikely explanation, because very large planetesimals would have to undergo
continuous collisional evolution in sufficient numbers. For the other older
system, ${\sim} \qty{200}{Myr}$ of age, this explanation can be strictly
excluded for the same reason: planetesimals would have to be unrealistically
large. For those older systems, a recent production of the planetesimal belt,
for instance, by a giant impact that could be associated with the late stage of
terrestrial planet formation, is a better option. Such giant impacts could have
occurred only several to tens of million years ago.

However, the true location of the planetesimal belts could be at larger
distances from the star than the mid-infrared flux center of the dust
distribution if the dust partly drifted inward under the Poynting–Robertson
effect in sufficient numbers without getting destroyed by collisions. In this
case, a primordial planetesimal belt could be at the origin of the warm dust
distribution in two ${\sim} \num{200}$ and ${\sim} \qty{400}{Myr}$ old A-type
systems. Thus, a detailed characterization of individual older systems is
necessary if one is to investigate whether a warm dust distribution can
originate from a planetesimal belt formed in the protoplanetary phase or a
recent giant collision.


\section*{Acknowledgments}

The authors thank the anonymous referee for their comments and suggestions,
which helped to improve the article. T.A.S. conducted parts of this work while
being an International Research Fellow of the Japanese Society for the
Promotion of Science (JSPS). The JSPS supported H.K. by Grants-in-Aid for
Scientific Research (JP21K03642, JP22H01278, JP22H00179, JP24K00690,
JP25K01055) from MEXT of Japan. T.A.S. and V.F.-G. acknowledge financial
support from the National Aeronautics and Space Administration (NASA) through
grants 80NSSC23K1473 and 80NSSC23K0288, and V.F.-G. in addition through grant
80NSSC21K0394. S.W. acknowledges the German Research Foundation (DFG) for
financial support under grant WO 857/15-2.


\software{
    \texttt{astropy} \citep{astropy_collaboration:2013,
    astropy_collaboration:2018, astropy_collaboration:2022},
    \texttt{Ipython} \citep{perez:granger:2007},
    \texttt{Jupyter} notebooks \citep{kluyver:2016},
    \texttt{Matplotlib} \citep{hunter:2007},
    NASA's Astrophysics Data System Bibliographic Services,
    \texttt{NumPy} \citep{harris:2020},
    \texttt{python} (\url{https://www.python.org/}),
    \texttt{SIMBAD} Astronomical Database \citep{wenger:2000},
}


\appendix


\section{Depletion Formula}
\label{sect_depletion_timescale}

For a collisional cascade, the mass distribution follows a power law with its
index depending on the size dependence of $Q\sbs{D}^\star$
\citep{kobayashi:tanaka:2010}. For the object sizes considered in this study,
$Q\sbs{D}^\star$ is dominated by the first two terms of eq.~\ref{eq_qds}, and
thus two power laws determine the mass distribution, which we model as
\begin{equation}
    n\sbs{s}(m) = \left\{
    \begin{array}{ll}
    A m^\alpha & {\rm for} \, m > m\sbs{tr}\\
    A^\prime m^{\alpha^\prime} & {\rm for}\, m < m\sbs{tr} \, ,
    \end{array}
    \right.
    \label{eq_power}
\end{equation}
where 
\begin{equation}
    \alpha = \frac{11+1.4}{6+1.4},\
    \alpha^\prime = \frac{11-0.3}{6-0.3}
\end{equation}
and 
\begin{equation}
    A m\sbs{tr}^\alpha = A^\prime m\sbs{tr}^{\alpha^\prime} \, .
    \label{eq_A_Ap}
\end{equation}
The transitional mass $m\sbs{tr}$ is assumed to be the mass equating the first
and second terms of $Q\sbs{D}^\star$, which gives  
\begin{equation}
        2.0 \times 10^7 \left( \frac{a\sbs{tr}}{\qty{1}{\cm}}\right)^{-0.3}  
        = 1.1 \left( \frac{a\sbs{tr}}{\qty{1}{\cm}}\right)^{1.4},
\label{eq_mtra}
\end{equation}
where $a\sbs{tr}$ is the radius of bodies with mass $m\sbs{tr}$. The surface
density $\Sigma$ and total optical depth $\tau$ of the debris disk are given by 
\begin{equation}
    \Sigma = \int m n\sbs{s}(m) \textrm{d} m \approx
               \frac{A m\sbs{pb}^{2-\alpha}}{2-\alpha},
    \label{eq_sigma_appro}
\end{equation}
\begin{equation}
    \tau = \int \pi a^2 n\sbs{s}(m) \textrm{d} m \approx
           \frac{A^\prime \pi a\sbs{bo}^2 m\sbs{bo}^{1-\alpha^\prime}}{
                \alpha^\prime - 5/3
           } \, ,
    \label{eq_tau_appro}
\end{equation}
where $m\sbs{bo}$ and $m\sbs{pb}$ are the masses corresponding to $a\sbs{bo}$
and $a\sbs{pb}$, respectively.

Since $\tau$ is proportional to $\Sigma$ (see eqs.~\ref{eq_sigma_appro} and
\ref{eq_tau_appro} with eq.~\ref{eq_A_Ap}), the formula for the time evolution
of $\Sigma$ in \citet{kobayashi:tanaka:2010} is modified to that for $\tau$,
which gives eq.~(\ref{eq_mass_loss}). The depletion timescale of $\tau$
corresponds to that of $\Sigma$ as given in \citet{kobayashi:tanaka:2010},
\begin{equation}
    t\sbs{dep} = \frac{m\sbs{pb}}{
                    (2-\alpha)^2 \Sigma \pi a\sbs{pb}^2 \xi \Omega
                } 
    \left( \frac{v^2}{2 Q\sbs{D}^\star} \right)^{-\alpha+1} s_{123} \, ,
    \label{eq_tdep_koba}
\end{equation}
where $s_{123} \approx 32.3$ is the depletion factor derived from the
integration of the mass distribution. Substituting
eqs.~(\ref{eq_power})–(\ref{eq_tau_appro}) into eq.~(\ref{eq_tdep_koba}) yields
\begin{equation}
    t\sbs{dep} = \frac{
                    a\sbs{pb}^{3\alpha-5}a\sbs{bo}^{-3 \alpha^\prime+5}
                    a\sbs{tr}^{3(\alpha^\prime-\alpha)}
                }{(2-\alpha)(\alpha^\prime-5/3) \xi \tau \Omega} 
    \left(\frac{v^2}{2 Q\sbs{D}^\star}\right)^{-\alpha+1} s_{123} \, . 
    \label{eq_tdep_app}
\end{equation}
Equation~(\ref{eq_dep_time}) is calculated from eq.~(\ref{eq_tdep_app}) using
$v = \sqrt{e^2+i^2}R\sbs{disk} \Omega$ with $e=2i=0.2$ and assuming that
$Q\sbs{D}^\star$ is dominated by its second term (see eq.~\ref{eq_qds}).


\section{Determination of the Blowout Grain Radius}
\label{appdx_sect_blowout_radius}

We assume the dust grains to be compact spheres of grain radius $a$ and compute
their absorption and scattering cross sections and their scattering phase
function via Mie theory \citep{mie:1908} using the numerical tool \texttt{miex}
\citep{wolf:voshchinnikov:2004}.

To compute the grain size below which dust grains are repelled from the system,
the combined force caused by radiation pressure and gravity has to be
evaluated. The radial component of the force $F\sbs{rp}$ caused by the pressure
of the stellar radiation field acting on spherical dust grains can be described
by \citep{kimura:1997}
\begin{equation}\label{eq_rad_pressure_force}
    F\sbs{rp} = \frac{\pi^2 a^2}{c} \left( \frac{R_\star}{r} \right)^2
    \int_0^{\infty} Q\sbs{rp} I_\lambda \mathrm{d} \lambda \, ,
\end{equation}
with the velocity of light in vacuum $c$, the intensity $I_\lambda$, the
stellocentric distance of the grain $r$, and the wavelength- and
grain-radius-dependent radiation pressure efficiency $Q\sbs{rp}$ derived from
Mie theory. To model the stellar intensity $I_\lambda$ we utilize the emission
spectrum of a blackbody with its temperature set to the stellar effective
temperature $T\sbs{eff}$.
Against the radiation pressure force acts the inward-directed gravitational
force
\begin{equation}
    F\sbs{G} = - G \frac{M_\star m}{r^2}
             = - G \frac{M_\star}{r^2} \frac{4\pi \rho a^3}{3} \, .
    \label{eq_grav_force}
\end{equation}
One defines the dimensionless $\beta$-ratio as the ratio of the two forces
\citep{zook:berg:1975} and obtains with eqs.~\ref{eq_rad_pressure_force} and
\ref{eq_grav_force}
\begin{equation}
    \beta = \frac{\lvert F\sbs{rp} \rvert}{\lvert F\sbs{G} \rvert}
          = \frac{3\pi}{4Gc} \frac{R_\star^2}{M_\star} \frac{1}{\rho a}
          \int_0^{\infty} Q\sbs{rp} I_\lambda \textrm{d} \lambda \, .
    \label{eq_beta_ratio}
\end{equation}
Under the assumption of the grains being released from a circular orbit, grains
with a $\beta$-ratio of $\beta \geq \num{0.5}$ will become unbound and leave
the system \citep{burns:1979, krivov:2006}. We compute the $\beta$-ratio using
eq.~\ref{eq_beta_ratio} for a range of grain radii and define the blowout
grain radius $a\sbs{bo}$ to be the radius for which $\beta = \num{0.5}$. While
the condition $\beta = 0.5$ can be true for more than one grain radius, in the
case of the considered luminous A-type stars, the solution is unique.


\section{Supplemental Figures}

\begin{figure*}
    \centering
    \includegraphics[width=1\linewidth]{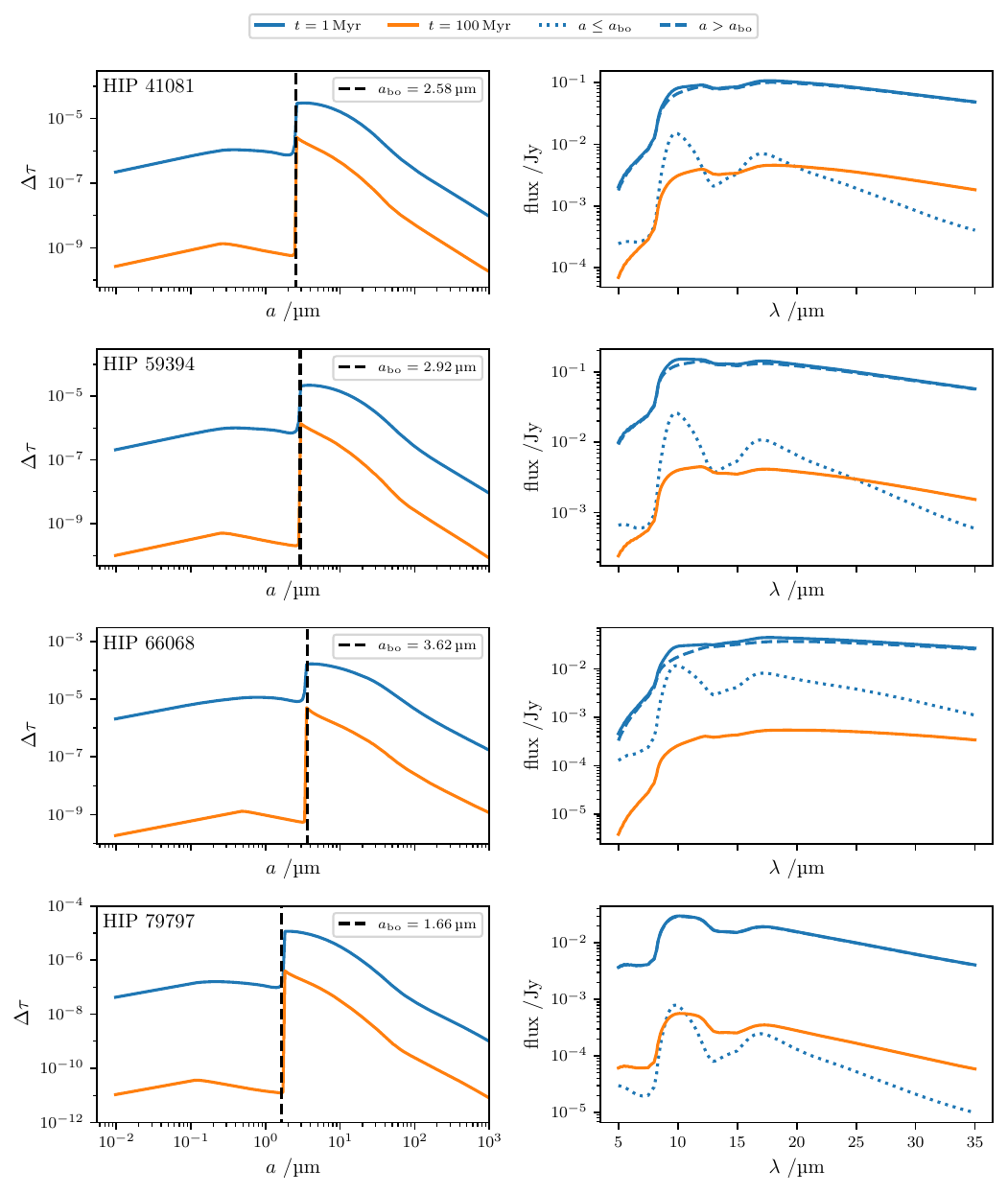}
    \caption{Same as Fig.~\ref{fig_opt_depth_sed}, but for all other systems.}
    \label{fig_multi_opt_depth_sed}
\end{figure*}

Figure~\ref{fig_multi_opt_depth_sed} shows the simulated size distribution of
optical depth and the corresponding SED at times of $t = \num{1}$ and
\qty{100}{Myr} for the systems HIP 41081, HIP 59394, HIP 66068, and HIP 79797.

\section*{ORCID iDs}

{\small
\noindent T. A. Stuber \orcidlink{0000-0003-2185-0525}%
\href{https://orcid.org/0000-0003-2185-0525}
     {https://orcid.org/0000-0003-2185-0525}\\
H. Kobayashi \orcidlink{0000-0001-8808-2132}%
\href{https://orcid.org/0000-0001-8808-2132}
     {https://orcid.org/0000-0001-8808-2132}\\
S. Wolf \orcidlink{0000-0001-7841-3452}%
\href{https://orcid.org/0000-0001-7841-3452}
     {https://orcid.org/0000-0001-7841-3452}\\
\mbox{V. Faramaz-Gorka \orcidlink{0000-0001-6403-841X}%
\href{https://orcid.org/0000-0001-6403-841X}
     {https://orcid.org/0000-0001-6403-841X}}
}


\clearpage
\bibliography{bibliography}{}
\bibliographystyle{aasjournal}


\end{document}